\documentclass[aps,prl,twocolumn,amsmath,amssymb,showpacs,superscriptaddress,notitlepage,longbibliography]{revtex4-1}
\usepackage{graphicx}
\usepackage{subfigure}
\usepackage{dcolumn}
\usepackage{bm}
\usepackage{amsfonts}
\usepackage{amssymb}
\usepackage{amsmath}
\usepackage{color}
\usepackage[colorlinks=true,breaklinks=true,linkcolor=blue,anchorcolor=red,citecolor=blue,urlcolor=blue]{hyperref}
\usepackage{longtable}
\usepackage{booktabs}
\usepackage{multirow}
\allowdisplaybreaks[3]

\def\bea{\begin{eqnarray}}
\def\eea{\end{eqnarray}}

\def\nn{\nonumber}
\def\Eq#1{Eq.~(\ref{#1})}

\def\Fig#1{Fig.~\ref{#1}}
\def\abs#1{\left|#1\right|}
\def\xk#1{\left(#1\right)}
\def\zk#1{\left[#1\right]}
\def\dk#1{\left\{#1\right\}}

\def\avg#1{\left\langle#1\right\rangle}

\def\pa{\partial}
\newcommand{\s}{{\sigma}}

\newcommand{\de}{\delta}
\newcommand{\De}{\Delta}
\newcommand{\ep}{\epsilon}
\newcommand{\ga}{\gamma}
\newcommand{\Ga}{\Gamma}
\newcommand{\ka}{\kappa}

\newcommand{\La}{\Lambda}

\renewcommand{\v}[1]{{\bf #1}}

\begin{document}

\title{Coulomb instabilities of 3D higher-order topological insulators}

\author{Peng-Lu Zhao}
\affiliation{Shenzhen Institute for Quantum Science and Engineering and Department of Physics, Southern University of Science and Technology (SUSTech), Shenzhen 518055, China}
\affiliation{Shenzhen Key Laboratory of Quantum Science and Engineering, Shenzhen 518055, China}

\author{Xiao-Bin Qiang}
\affiliation{Shenzhen Institute for Quantum Science and Engineering and Department of Physics, Southern University of Science and Technology (SUSTech), Shenzhen 518055, China}
\affiliation{Shenzhen Key Laboratory of Quantum Science and Engineering, Shenzhen 518055, China}

\author{Hai-Zhou Lu}
\email{Corresponding author: luhz@sustech.edu.cn}
\affiliation{Shenzhen Institute for Quantum Science and Engineering and Department of Physics, Southern University of Science and Technology (SUSTech), Shenzhen 518055, China}
\affiliation{Shenzhen Key Laboratory of Quantum Science and Engineering, Shenzhen 518055, China}

\author{X. C. Xie}
\affiliation{International Center for Quantum Materials, School of Physics, Peking University, Beijing 100871, China}
\affiliation{CAS Center for Excellence in Topological Quantum Computation, University of Chinese Academy of Sciences, Beijing 100190, China}
\affiliation{Beijing Academy of Quantum Information Sciences, West Building 3, No. 10, Xibeiwang East Road, Haidian District, Beijing 100193, China}

\date{\today}

\begin{abstract}
Topological insulator (TI) is an exciting discovery because of its robustness against disorder and interactions. Recently, higher-order TIs have been attracting increasing attention, because they host 1D topologically-protected hinge states in 3D or 0D corner states in 2D.
A significantly critical issue is whether the higher-order TIs also survive interactions, but it is still unexplored. We study the effects of weak Coulomb interaction on a 3D second-order TI, with the help of a renormalization group calculation. We find that the 3D higher-order TIs are always unstable, suffering from two types of topological phase transitions. One is from  higher-order TI to TI, the other is to normal insulator (NI). The first type is accompanied by emergent time-reversal and inversion symmetries and has a dynamical critical exponent $\kappa=1$. The second type does not have the emergent symmetries and has non-universal dynamical critical exponents $\kappa<1$. Our results may inspire more inspections on the stability of higher-order topological states of matter and related novel quantum criticalities.
\end{abstract}

\maketitle

\begin{figure}[htbp]
\center
\includegraphics[width=0.5\textwidth]{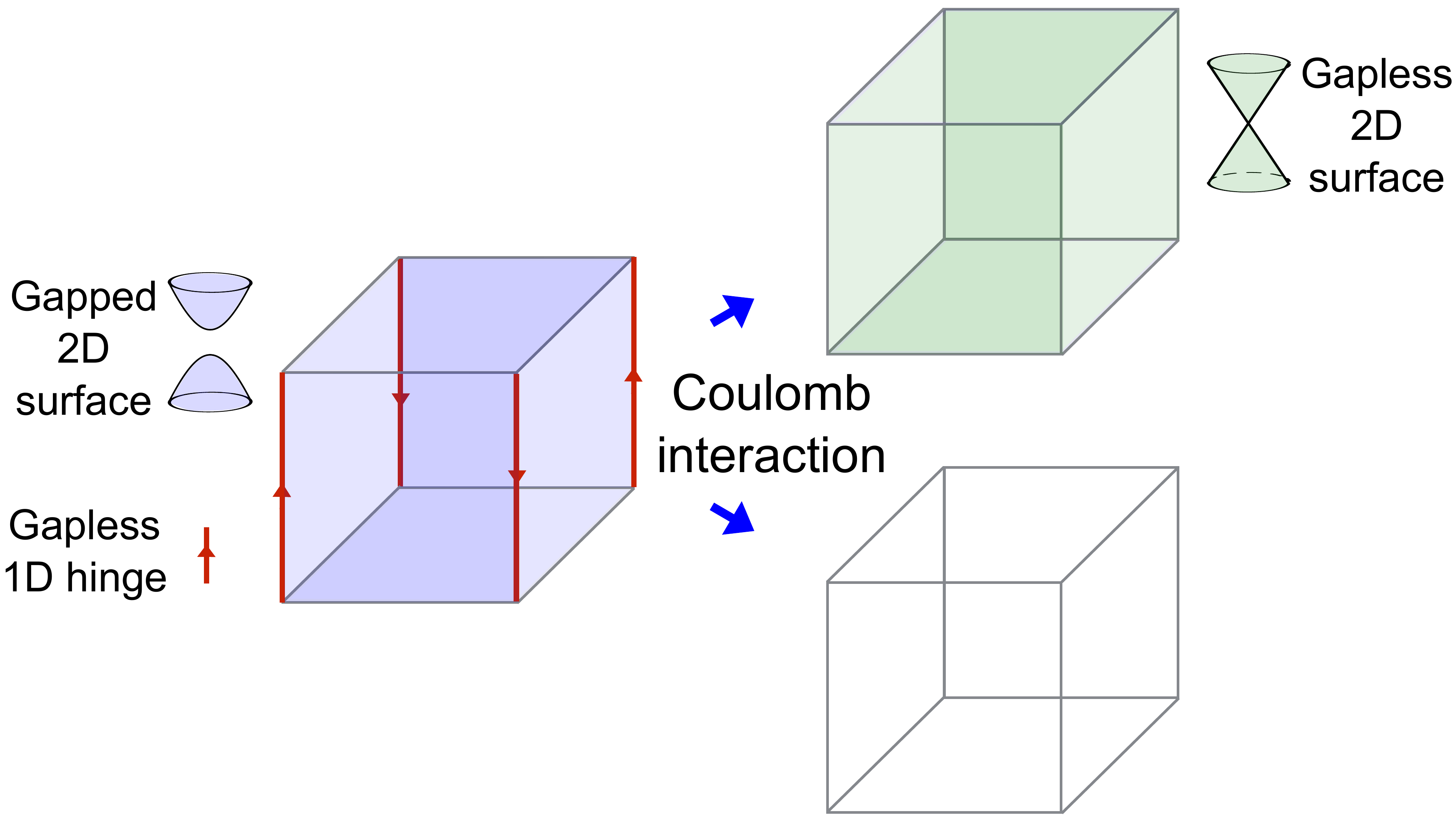}
\caption{Schematic of a typical 3D second-order topological insulator (TI) (left) which hosts 3D gapped bulk states in the interior but topologically protected 2D massive (gapped) Dirac cones on the surfaces and gapless 1D chiral hinge states. It may turn to a TI (right top) or a normal insulator (NI) (right bottom), in the presence of the Coulomb interaction.} \label{FHOTI}
\end{figure}

\textcolor[rgb]{0.00,0.00,1.00}{\emph{Introduction}.}--As generalizations of the topological insulator (TI) \cite{Ful07L,Moore07B,Murakami07NJP,RoyR09B,Ful07B,Hsieh08N,
Zhanghj09NP,Xiay09NP,Shen17b,Hasan10RMP,Qixl11RMP,Hasan11AR}, higher-order TIs have been attracting considerable interest recently \cite{Benalcazar17S,Benalcazar17B,Songzd17L,Wangcm17L,Langbehn17L,
Schindler18SA,Schindler18NP,Ezawa18L,Lizx19npj,
Liut19L,Luoxw19L,Kudo19L,Chenr20L,Agarwala20R,Fu21arXiv}. A simplest 3D higher-order TI hosts 3D gapped bulk states inside but topologically protected gapless 1D hinge states and gapped 2D surface states (Fig. \ref{FHOTI}). There have been experimental evidences for the higher-order topology in bosonic systems, including circuitry  \cite{Peterson18N,Imhof18NP,Serra-Garcia119B}, phononics \cite{Serra-Garcia18N}, acoustics \cite{Xuehr19NM,Nix19NM,Xuehr20NC}, and photonics \cite{Mittal19NPO,Hassan19NPO,Xieby19L,Limy20NPO}.
Despite the theoretical predictions on material candidates  \cite{Schindler18SA,Schindler18NP, Yuecm19NP,Wangzj19L,Xuyf19L,Zhangrx20L}, there are few observation of higher-order TIs in electronic systems \cite{Schindler18NP}.
This raises the concerns on the stability of higher-order TIs against, \emph{e.g.} disorder \cite{Suzx19CPB,Wangc20R,Lica20L,Szabo20R}.
More importantly, it is still unknown whether higher-order TIs can survive a more intrinsic presence in electronic systems, the Coulomb interaction \cite{Gonzalez99B,Goswami11L,Hosur12L,Moon13L,Yang14NP,Hofmann14L,Laihh15B,Jiansk15B,Huh16B,Isobe16L,Cho16SR,Isobe16LII,Szabo20R}.

In this Letter, we study the stability of 3D higher-order TIs in the present of the Coulomb interaction. We find that the higher-order TIs are always unstable. Two types of topological phase transition could happen (Fig. \ref{FHOTI}). In the first type, the Coulomb interaction could drive a topological phase transition from higher-order TI to TI. This transition is accompanied by the recovery of time-reversal and inversion symmetries in the low-energy limit, and hence is protected by these emergent symmetries. The quantum criticality for this phase transition is described by a dynamical critical exponent $\kappa=1$ and a correlation length exponent $\nu=1$. In the second type, the Coulomb interaction could induce another topological phase transition from higher-order TI to a normal insulator (NI). There are no emergent symmetry when this phase transition happens, and its criticality is characterized by the appearance of non-universal dynamical critical exponents $\kappa<1$. Our results will be insightful for the ongoing experimental search for higher-order TIs in electronic systems.

\textcolor[rgb]{0.00,0.00,1.00}{\emph{Model for 3D second-order TIs}.}--We start with a four-band Hamiltonian for 3D second-order TIs \cite{Schindler18SA},
\bea
\mathcal{H}_{\mathrm{0}}(\v{k}) &=& \big[M+\sum_{i}t_i \cos (a k_i) \big]  \tau_z \sigma_0 +\sum_{i} \Delta_i\sin \xk{a k_i} \nn\\&&\times\tau_x \sigma_i
+\Delta_2 \zk{\cos (a k_x) - \cos (a k_y) } \, \tau_y \sigma_0,\label{EqHam}
\eea
where $i=x,y,z$, $a$ is the lattice constant and $\sigma_i$ and $\tau_i$ are the Pauli matrices. $M, t_i, \De_i, \De_2$ are the hopping parameters, and we take $t_x=t_y=t_{\perp}$, and $\Delta_x=\Delta_y=\Delta_{\perp}$. This model has no time-reversal symmetry if $\Delta_2 \neq 0$ (SI of \cite{Supp}).
Also, a four-fold rotation symmetry
%on the $x-y$ plane by the operator
$R_{4z} \equiv \tau_0 e^{-\mathrm{i} \frac{\pi}{4} \sigma_z}$ is broken by the nonzero $\Delta_2$ term.
The combination $R_{4z}\mathcal{T}$ ($\mathcal{T}$  is the time-reversal operator) is a symmetry %of Hamiltonian, and this symmetry
that protects
%the existence of
the 3D second-order TI.
Another important symmetry is the combination of time reversal and inversion $\mathcal{IT}$ (SII of \cite{Supp}), where $\mathcal{I}=\tau_z\s_0$.
With which,
%the $\mathcal{IT}$ symmetry, one can identify
the $\mathcal{Z}_2$ invariants
%are identifies by
\bea
(-1)^{\vartheta}=\prod_{i} \prod_{n=1}^{N / 2} \xi_{n}\xk{\Ga_i}, \label{EqToin}
\eea
where $\xi_{n}\xk{\Ga_i}=\pm 1$ is the eigenvalue of $\mathcal{I}$ for the $n$th occupied energy band at momenta $\Ga_i$, and $\Ga_i
\in \{(0,0,0),(\pi, \pi, 0),(0,0, \pi),(\pi, \pi, \pi)\}$ representing all the $R_{4z} \mathcal{T}$-invariant $\v k$ points. As a result, for $\abs{2t_{\perp}-t_z}<\abs{M}<\abs{2t_{\perp}+t_z}$, $(-1)^{\vartheta}=-1$ (SII of \cite{Supp}), which represents the second-order TI and for $\abs{M}>\abs{2t_{\perp}+t_z}$ or $\abs{M}<\abs{2t_{\perp}-t_z}$, $(-1)^{\vartheta}=1$, which stands for a NI.
%the topologically-trivial band insulator phase.
This difference establishes only when $\De_i \neq 0 \neq \De_2$. Once $\De_i =0$, there exist gapless points which break the insulating nature. Once $\De_2=0$, time-reversal symmetry recovers and
the
%topological nontrivial
phase is a TI.
Later, we will show that, even if starting with $\De_2\neq 0$ and $\abs{2t_{\perp}-t_z}<\abs{M}<\abs{2t_{\perp}+t_z}$,  the Coulomb interaction makes
$\De_2$ flow to zero in the low-energy limit, leading to a transition from higher-order TI to TI, or causes $\abs{M}>\abs{2t_{\perp}+t_z}$, which induces a transition from higher-order TI to NI.

\textcolor[rgb]{0.00,0.00,1.00}{\emph{Coulomb interaction and renormalization group equations.}}--The
%low-energy
effective action in Euclidean spacetime for the second-order TI in the presence of the Coulomb interaction takes the form
\bea
\mathcal{S}&=&\int d\tau d^3 \v r \big\{ \bar{\psi}\big[ \xk{\pa_{\tau}+ig\phi}\ga_0+v_i\ga_i\pa_i+m+B_i\pa_i^2
\nn\\&&
-i D \xk{\pa_x^2 - \pa_y^2} \ga_5  \big]\psi
+ \frac{1}{2}\eta_i\left(\partial_{i} \phi\right)^{2}\big\}, \label{EqCoulAct}
\eea
where $\psi$ describes a four-component fermion field and $\bar{\psi}=\psi^{\dagger}\ga_0$. The $\ga$ matrices satisfy the anticommuting algebra $\dk{\ga_{\mu},\ga_{\nu}}=2\de_{\mu,\nu}$. The repeated index $i$ sums
%automatically
for $i=x,y,z$, and $v_{i}=\Delta_{i}a$, $m=M+2t_{\perp}+t_z$, $B_i=t_ia^2/2$, $D=\Delta_2a^2/2$, which are obtained by expanding \Eq{EqHam} around the $\Ga$ point.
The parameter
%$v_i$ represents the Fermi velocities,
$m$ is the Dirac mass, and the $B_i$ and $D$ terms represent the quadratic corrections to the Dirac Hamiltonian.
We introduce an auxiliary scale field $\phi$ through the Hubbard-Stratonovich transformation \cite{Colemanb} to decouple the density-density Coulomb interaction. $\xk{\eta_x,\eta_y,\eta_z}=\xk{1,1,\eta}$ is used to characterize the spatial anisotropy of $\phi$. $g=e/\sqrt{\ep}$ representing the coupling between electrons and the scalar field, where $-e$ is the electron charge and $\epsilon$ is the dielectric constant.
The Coulomb interaction does not break the $R_{4z}\mathcal{T}$ and $\mathcal{IT}$ symmetries of \Eq{EqHam} (SIIIA of \cite{Supp}), and the topology is still distinguished by \Eq{EqToin}. According to \Eq{EqHam}, $m$ solely controls the gap closing and reopening \cite{Murakami07NJP,Shen11spin,Shen17b}, and hence its sign change identifies the phase transition between higher-order TI and NI.

To explore how the Coulomb interaction renormalizes the parameters and consequently leads to the phase transitions, we perform a Wilsonian momentum-shell renomalization group analysis \cite{Shankar94RMP} for
%the model described by
\Eq{EqCoulAct}. After redefining the original parameters $B_i, D, m, v_i$, and Coulomb interaction strength $g$ into dimensionless
\begin{eqnarray}
B_i\La v^{-1}\eta_i^{-1}\rightarrow B_i, \,\, D\La v^{-1}\rightarrow D,
\nn \\
mv^{-1}\La^{-1}\rightarrow m, \,\, g^{2}/(4\pi^2v\sqrt{\eta})\rightarrow \alpha,
\end{eqnarray}
where $\La$ is the cutoff, the renormalization group flow equations
for them
%these dimensionless parameters
are found as (SIIIC of \cite{Supp})
\bea
d v/d\ell&=&\xk{\ka-1+\mathcal{F}_{0}^{\perp}\alpha}v,
\label{EqRGv}
\\
d \ga^2/d\ell&=&\ga^2  \alpha\zk{2\xk{2\mathcal{F}_{0}^{z}
-\mathcal{F}_{0}^{\perp} } -\mathcal{F}_{2}^{z}+\mathcal{F}_{2}^{\perp}},\label{EqRGga}
\\
d m/d\ell&=&m+\alpha  \zk{\xk{m-B_{z}}\mathcal{F}_{0}^{z}-B_{\perp}\mathcal{F}_{0}^{\perp}},
\label{EqRGm}
\\
d B_{\perp}/d\ell&=&
-B_{\perp}+\alpha \Big[B_{\perp} \xk{\mathcal{F}_{1}^{\perp}-\mathcal{F}_{0}^{\perp}}
\nn\\&&+\xk{B_z-m}\mathcal{F}_{1}^{z}
-m\mathcal{F}_{1}^{\perp} \Big] ,\label{EqRGBperp}
\\
d B_{z}/d\ell&=&
-B_{z}+\alpha  \Big[B_{z}\xk{\mathcal{F}_{2}^{\perp}-\mathcal{F}_{2}^{z}
-\mathcal{F}_{1}^{\perp}-3\mathcal{F}_{1}^{z}} \nn\\&&
+B_{\perp}\xk{\mathcal{F}_{0}^{\perp}-2\mathcal{F}_{1}^{\perp}}-m\xk{3\mathcal{F}_{0}^{z}-\mathcal{F}_{0}^{\perp}} \Big],\label{EqRGBz}
\\
d D/d\ell&=&\zk{-1+\alpha \xk{\mathcal{F}_{1}^{D}-\mathcal{F}_{0}^{\perp} }}D,\label{EqRGD}
\\
d\alpha/d\ell&=&-\alpha^2\zk{\mathcal{F}_{0}^{\perp}+\xk{\mathcal{F}_{2}^{z}+\mathcal{F}_{2}^{\perp} }/2},\label{EqRGalpha}
\eea
where $v=v_{x,y}$, $\kappa$ is the dynamic exponent whose value is obtained by fixing $v$, $\ga^2=v_z/(v\eta)$ describing the anisotropy, and $\ell$ is the running scale parameter whose value increase lowers the energy scale. $\mathcal{F}_{0-2}^{\perp, z}, \mathcal{F}_{1}^{D}$ are dimensionless functions of $m$, $B_i$, $D$, and $\ga^2$ (Their expressions in IIIB of \cite{Supp}).
We numerically solve the
%above coupled
renormalization group equations and obtain the running of $m, B_i, D, \alpha, \ga^2$ and $\kappa$ with $\ell$. Despite that the running of these parameters highly depends on their initial values, \emph{i.e.} their values at the cutoff $\La$,
their behaviors can be classified into two categories, corresponding to two types of phase transitions.

\begin{figure}[htb]
\centering
\hspace{-0.2cm}
\includegraphics[width=3.3in]{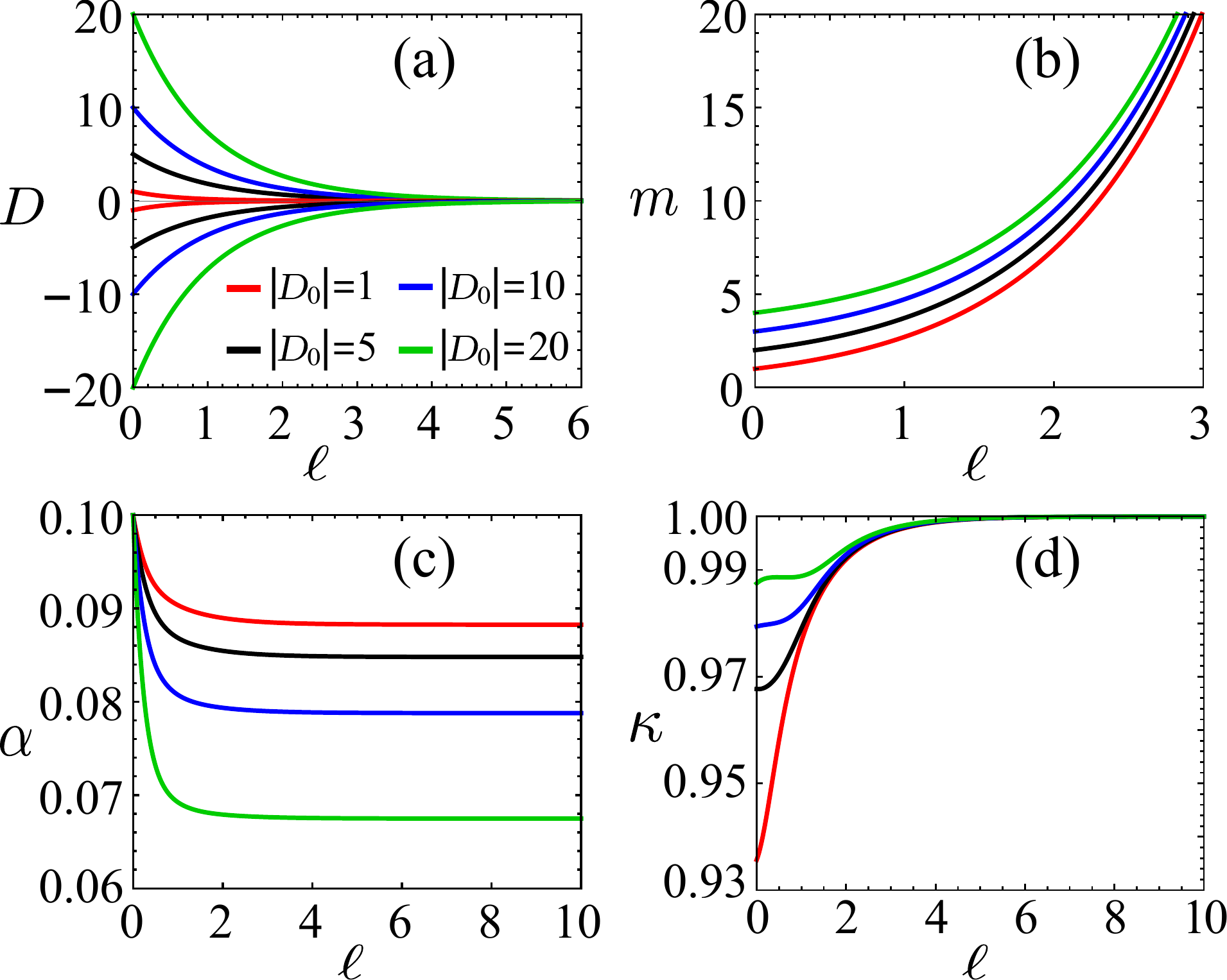}
\caption{[(a)--(c)]
%Numerical results of
The renormalized $D$, $m$, and $\alpha$ as functions of the running scale parameter $\ell$.
$D$ and $m$ protect the second-order and first-order topological properties, respectively. The vanishing $D$ and increasing $m$ means a topological phase transition from second-order TI to TI.
(d) The scale dependence of $\kappa$. The solutions are obtained by fixing
%the values at $\ell=0$ as
the initial values $m_0=B_{\perp}^{0}=1$, $B_{z}^{0}=0.5$, $\alpha_0=0.1=\ga^2_0=0.1$ while varying $D_0$.
The same legends in (a)--(d). In (b), the
$m\xk{\ell}$ curve of $D_0=5, 10$, and $20$ are shifted
vertically by $1$ for clarity.}\label{FSotoTI}
\end{figure}

\begin{figure}[htb]
\centering
\includegraphics[width=3.3in]{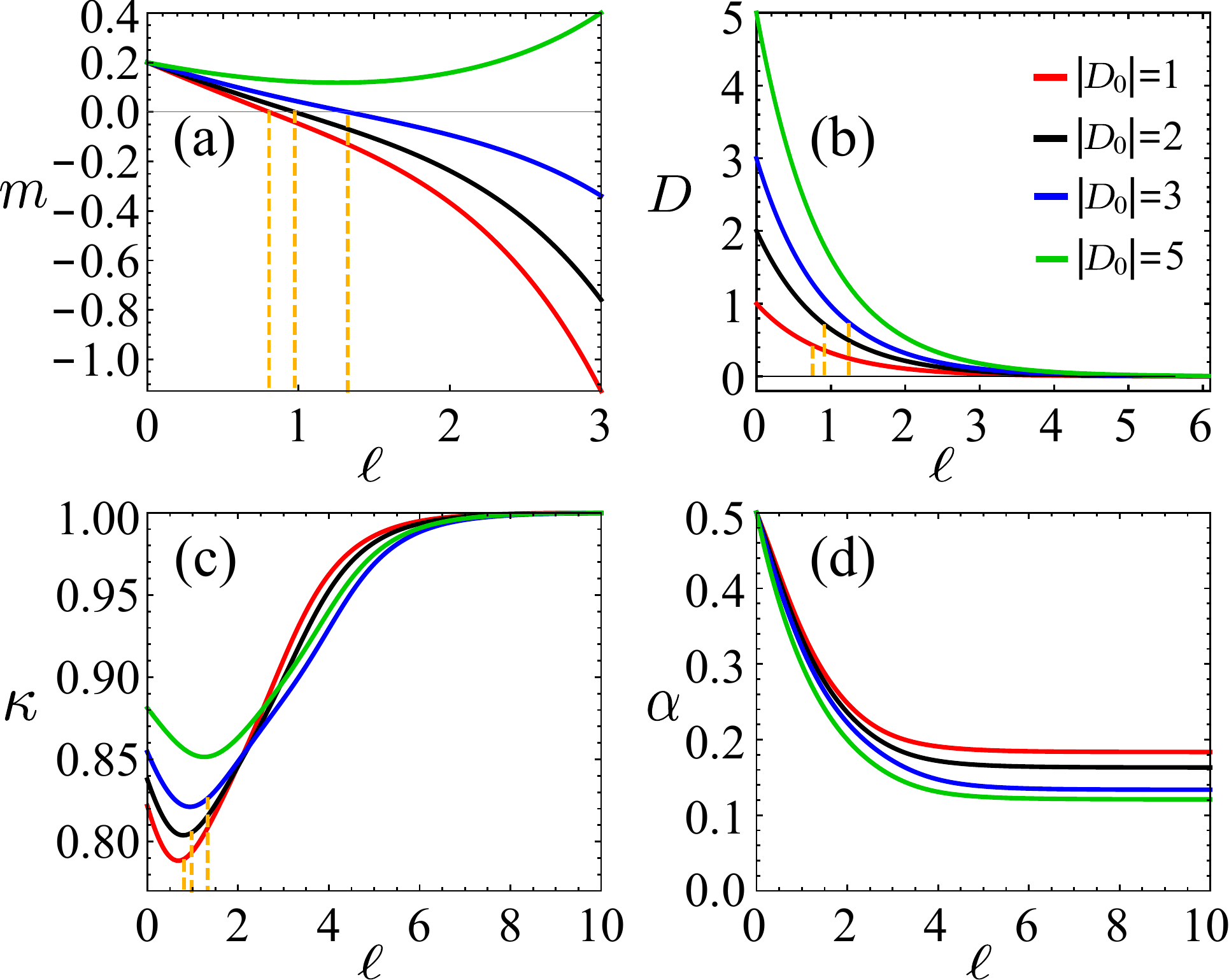}
\caption{[(a), (b), and (d)]
%Numerical results of
The renormalized $m$, $D$, and $\alpha$ as functions of the running scale parameter $\ell$. $D$ and $m$ protect the second-order and first-order topological properties, respectively. The finite $D$ at the sign change of $m$ means a topological phase transition from 3D second-order TI to NI.
(c) The scale dependence of $\kappa$. The solutions are obtained by varying the initial value of $D$ while fixing the initial values of other parameters as $m_0=0.2$, $B_{\perp}^0=2B_z^0=2$, $\alpha_0=0.5$, and $\ga^2_0=0.1$. The same legends in (a)--(d).
The orange dashed lines in (a)--(c) label the values of $\ell$ at the phase transition points. In (b) and (c), the crossing points of the orange dashed lines with the red, black, blue lines indicate the values of $D$ and $\kappa$ at the topological phase transition points. From left to right, $D=0.43, 0.72, 0.74$ in (b)  and $\kappa=0.79, 0.81, 0.83$ in (c).}\label{FtoBI}
\end{figure}

\textcolor[rgb]{0.00,0.00,1.00}{\emph{From second-order TI to TI.}}--This phase transition is characterized by a vanishing $D$ and always positive $m$ at large $\ell$ (the low-energy limit), as  shown in \Fig{FSotoTI}.
Figure \ref{FSotoTI}(a) shows that, in a large range of $D_0$, $D$ flows to zero rapidly with increasing $\ell$. This behavior reflects that the renormalization group equation for $D$ [\Eq{EqRGD}] only has one stable fixed point at $D_{\ast}=0$. As $D$ flows to zero, $m$ increases and remains positive [\Fig{FSotoTI}(b)]. The unrestricted growth of $m$ ceases the rapid decay of $\alpha$ [\Fig{FSotoTI}(c)] because $\mathcal{F}_{0}^{\perp}\sim 1/|m|, \mathcal{F}_{2}^{\perp,z}\sim 1/|m|^3$ for extremely large $m$, which give rises to $d \alpha/d\ell \sim 0$. Although the effective Coulomb interaction is marginally irrelevant according to \Eq{EqRGalpha}, its values in the low-energy limit is a small constant instead of zero [see \Fig{FSotoTI}(c)].
With no sign change of $m$, there is no gap closing and reopening near the $\Ga$ point when approaching the low-energy limit, and the topological invariant does not change. However, when $D$ flows to zero, the free part of the effective model \Eq{EqCoulAct} reduces to the modified Dirac Hamiltonian that describes TIs \cite{Zhanghj09NP,Shen17b}. According to the previous results \cite{Qixl08B,Wangz10L,Goswami11L,Wangz12B}, TIs are immune to weak Coulomb interactions. Therefore, the low-energy state is a TI with a finite but weak Coulomb interaction, which means that the second-order TI is unstable due to the Coulomb interaction.
Time-reversal symmetry and inversion symmetry emerge along with the phase transition from second-order TI to TI. The previous works have shown the possibilities of emergent Lorentz \cite{Nielsen78NPB,Chadha83NPB,Sitte09L,Anber11D,Bednik13JHEP,Sibiryakov14L,Kharuk16TMH,Roy16JHEP}, chiral \cite{Balachandran97IJ,Candido18B,Szabo21JHEP}, and super- \cite{Balents98I,Fendley03L,Lee07B,Yuy08L,Yuy10L,Bauer13B,Grover14S, Ponte14NJP,Jiansk15L,Jiansk17L,Feldmann18D,Zhaopl19npj} symmetries.
Our concrete example above shows the emergent discrete time-reversal and inversion symmetries, enriching the family of emergent symmetries \cite{Volovik08P}.
This phase transition does not need a large critical value of $\alpha$, and exists at least for $\alpha \sim 10^{-3}$, corresponding to an extremely weak Coulomb interaction (IV of \cite{Supp}).
We find that the dynamical critical exponent $\kappa=1$ for this phase transition, as shown in \Fig{FSotoTI}(d). We also obtain a correlation length exponent $\nu=1$ by assuming that the spatial correlation length $\xi$ diverges as $\delta \equiv D-D_{*} \rightarrow 0$ in the manner $\xi \sim |\de|^{-\nu}$. This definition is similar to the conventional definition of the correlation length exponent in a symmetry-broken quantum phase transition \cite{Sondhi97RMP,Imada98RMP}.

\textcolor[rgb]{0.00,0.00,1.00}{\emph{From second-order TI to NI.}}--This phase transition is characterized by a sign change of $m$ and a finite $D$ as $m$ changes sign, as shown in \Fig{FtoBI}. Here, the finite $D$
%means that it cannot been viewed as zero approximately, which
guarantees that the topological phase transition to NI happens before the transition to TI.
%We show the existence of this kind of behaviors for $m$ and $D$ .
Figure \ref{FtoBI}(a) shows that $m$ changes sign when $D_0$ is below a critical value. For comparison, the green line shows a case in which $m$ does not change sign.
Figure \ref{FtoBI}(b) shows that $D$ does not vanish when $m$ changes sign.
Therefore, for the parameters represented by the red, black, and blue lines in \Fig{FtoBI}(a), the transitions from second-order TI to NI happen and for the case depicted by the green line the transition from second-order TI to TI happens.
%, which has been discussed previously.
Due to the finite $D$, there is no emergent time-reversal and inversion symmetries at the phase transition point. After the transition, the emergent time-reversal and inversion symmetries appear in the low-energy limit of the NI.
Interestingly, this transition has no universal dynamical critical exponent.
Figure \ref{FtoBI}(c) shows that the dynamical critical exponent has different values for the three cases, all below $1$. A non-universal $\kappa<1$ is a direct result of the transition happening at a finite energy scale $\ell$.
According to \Eq{EqRGv},
\bea
\kappa\xk{\ell}=1-\alpha\xk{\ell}\mathcal{F}_{0}^{\perp}\xk{\ell}.
\eea
The values of $\alpha\xk{\ell}$ are always positive constant [see \Fig{FtoBI}(d)], and the non-negative $\mathcal{F}_{0}^{\perp}\xk{\ell}$ vanishes only as $\ell\rightarrow \infty$. Considering that the sign change of $m$ always happens at a finite $\ell$, the value of $\kappa$ is smaller than $1$ and its particular value depends on $\alpha$ and $\mathcal{F}_{0}^{\perp}\xk{\ell}$, and hence is non-universal. However, as shown in \Fig{FtoBI}(c), the asymptotic value of $\kappa$ as $\ell \rightarrow \infty $ is still $1$, which describes the low-energy dynamics of the NI.

We have shown our main results by varying the initial value of $D$ while fixing those of the other parameters.
We could perform similar analyses for any cases by changing the initial value of one parameter while fixing others, but our conclusion still holds.

\begin{figure}[htb]
\centering
\includegraphics[width=3.4in]{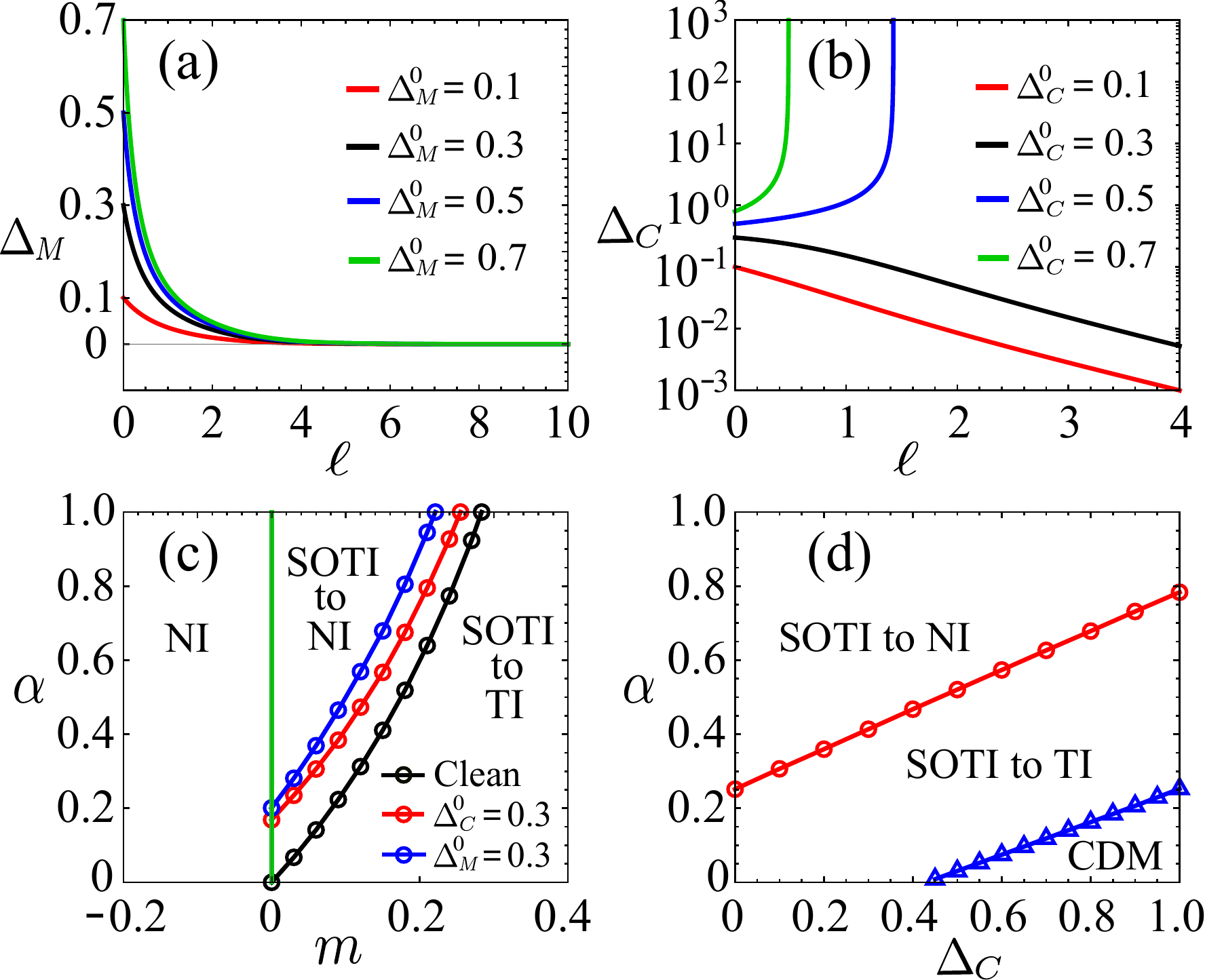}
\caption{[(a) and (b)]
%Numerical results of
The renormalized $\De_M$ and $\De_C$. The different curves are obtained by fixing the initial values $m_0=B_{\perp}^0=B_z^0=\ga^2_0=0.1$, $\alpha_0=0.5$ while varying the initial values of $\De_M$ in (a) and $\De_C$ in (b). [(c) and (d)] Phase diagrams in the $m$--$\alpha$ and $\De_C$--$\alpha$ planes. In (c), the black, red, and blue lines represent the boundary for the cases without disorder, with random mass $\De_M^{0}=0.3$, and with random chemical potential $\De_C^{0}=0.3$, respectively. In (d), CDM stands for the compressible diffusive metal. $D_0=B_{\perp}^0=B_z^0=1, \ga^2_0=0.1$ for the both diagrams and $m_0=0.1$ for (d). }\label{FPhadia}
\end{figure}

\textcolor[rgb]{0.00,0.00,1.00}{\emph{Screened Coulomb interaction.}}--The screening of Coulomb interaction cannot change our main conclusion. Due to the existence of gap $m$, once
the Fermi energy is placed in the gap, the density of state vanishes. The screening effect is extremely weak, compared to those in metals and semimetals. Using the random phase approximation, the renormalized boson propagator accounting for the screened Coulomb potential is
\bea
k_{\perp}^2+\eta k_z^2-\Pi\xk{k} &\sim& k_{\perp}^2\xk{1+ \alpha \mathcal{F}_{2}^{\perp}\ell}
\nn\\&+&\eta k_{z}^2\xk{1+ \alpha \mathcal{F}_{2}^{z}\ell}.
\eea
Considering $\mathcal{F}_{2}^{\perp,z}\sim 1/|m|^3 \sim 0$ at low energies, the screening effect therefore can be ignored.

\textcolor[rgb]{0.00,0.00,1.00}{\emph{Effect of coexisting disorder.}}--
%To consider disorder effects, w
We introduce disorder described by $\de H=U_i(x)\bar{\psi} \Ga_i \psi$, where $\Ga_i$ is a $4\times 4$ Hermitian matrix and $U_i(x)$ is the impurity potential of a Gaussian white-noise distribution as $\avg{U_i}=0$ and $\avg{U_i(\v x)U_j\xk{\v x'}} =\De_i\de_{ij} \de(\v x-\v x')$.
The types of disorder which respect $R_{4z}\mathcal{T}$ and $\mathcal{IT}$ symmetries are denoted by $\Ga_i=\mathcal{I}_{4\times4}$ and $\Ga_i=\ga_0$ (SVI of \cite{Supp}), and are referred to as the random mass and random chemical potential, respectively. The coupling strength $\De_M$ for the random mass is irrelevant [\Fig{FPhadia}(a)] and hence its existence cannot prevent the phase transitions in the clean system. As shown in \Fig{FPhadia}(b), the coupling strength $\De_C$ has a critical value $\De_C^{c}$ for the random chemical potential. Once the initial value $\De_C^{0}$ is smaller than $\De_C^{c}$, the random chemical potential is also irrelevant and cannot change our conclusion (SVI of \cite{Supp}). Due to the disorder-induced renormalization to the gap $m$, the boundary between the two kinds of phase transitions are shifted. As shown in \Fig{FPhadia}(c), the boundary between the phase transition to TI and NI for the clean system, random mass, and random chemical potential are different.
If $\De_C^{0}>\De_C^{c}$, a disorder-induced phase transition happens and the system flows to a disorder-dominated phase, dubbed the compressible diffusive metal \cite{Wangc20R,Supp}.

\textcolor[rgb]{0.00,0.00,1.00}{\emph{Phase diagrams.}}--To have a global view of the various phases, \Fig{FPhadia} shows two phase diagrams.
According to \Fig{FPhadia}(c), we find that $m$ plays a key role to determine the type of phase transition.
Once $m$ is large enough, the transition to NI cannot happen. Due to the dominant role of $m$, our conclusion also apply to other 3D higher-order TIs (e.g., the helical second-order TI \cite{Schindler18SA}), in particular when the topology depends on the quadratic or higher-order corrections to the Dirac Hamiltonian (e.g., the $D$-term in our model). According to our calculation, these terms are all irrelevant in the low-energy limit, which causes the transition to TI. The transition from higher-order TI to NI originates from the interplay between the anisotropic quadratic term ($B_i$) and Dirac mass term ($m$) generated by the Coulomb interaction, which does not depend on the terms protecting the higher-order TIs and hence will still exist for other 3D higher-order TIs.

\textcolor[rgb]{0.00,0.00,1.00}{\emph{Discussion}.}--Our results show that the Coulomb interaction is critical in the experiments searching for higher-order TIs. Previously, the transition between TI and NI has been observed in $\text{BiTl}\left(\text{S}_{1-\delta} \text{Se}_{\delta}\right)_{2}$ \cite{Xusy11S} by varying $\de$.
Our theory shares a similar spirit, by fixing the values of the parameters at the cutoff and analyzing their behaviors at a particular low-energy scale.
As a result, the change of initial values of the parameters is equivalent to the change of the parameters at the particular energy scale relevant to the experiment.
Therefore, the topological phase transitions from higher-order TI to NI or TI are possible in experiments by doping higher-order TIs, such as
the recently-proposed candidate materials bismuth \cite{Schindler18NP}, Eu$\text{In}_2\text{As}_2$ \cite{Xuyf19L}, and Mn$\text{Bi}_2\text{Te}_4$ \cite{Zhangrx20L}. Correspondingly, the 1D gapless hinge state will transform into a gapped state or 2D Dirac cone, which may be observed.

\begin{acknowledgements}
We thank helpful discussions with Jing-Rong Wang and Daoyuan Li. This work was supported by the National Natural Science Foundation of China (12047531, 11534001, 11925402), the Strategic Priority Research Program of Chinese Academy of Sciences (XDB28000000), the National Key R \& D Program (2016YFA0301700), Guangdong province (2016ZT06D348, 2020KCXTD001), Shenzhen High-level Special Fund (G02206304, G02206404), and the Science, Technology and Innovation Commission of Shenzhen Municipality (ZDSYS20170303165926217, JCYJ20170412152620376). The numerical calculations were supported by Center for Computational Science and Engineering of SUSTech.
\end{acknowledgements}

\bibliographystyle{apsrev4-1-etal-title_6authors}
\bibliography{HOTICoul0502}

\end{document}